\documentclass[showpacs,twocolumn]{revtex4}
\usepackage{graphicx}
\usepackage{color}
\usepackage{times}

\begin{document}
\title{Equipartition of rotational and translational energy in a dense granular gas}
\author{Kiri Nichol$^1$} 
\author{Karen E. Daniels$^2$} \email{kdaniel@ncsu.edu}
	\affiliation{$^1$Kamerlingh Onnes Laboratory, Leiden University, The Netherlands}
	\affiliation{$^2$Department of Physics, North Carolina State University, Raleigh, North Carolina, USA 27695}
\date{\today}

\begin{abstract}
Experiments quantifying the rotational and translational motion of particles in a dense, driven, 2D granular gas floating on an air table reveal that kinetic energy is divided equally between the two translational and one rotational degrees of freedom. This equipartition persists when the particle properties, confining pressure, packing density, or spatial ordering are changed. While the translational velocity distributions are the same for both large and small particles, the angular velocity distributions scale with the particle radius. The probability distributions of all particle velocities have approximately exponential tails.  Additionally, we find that the system can be described with a granular Boyle's Law with a van der Waals-like equation of state. These results demonstrate ways in which conventional statistical mechanics can unexpectedly apply to non-equilibrium systems.
\end{abstract}


\maketitle


Many processes operate far from equilibrium and therefore cannot be described within the framework of classical thermodynamics and statistical mechanics. One of the simplest non-equilibrium systems is the granular gas, composed of driven, colliding macroscopic particles \cite{Goldhirsch-2003-RGF}. Such a system differs from an ideal gas in two important ways: energy input is required to maintain a gaseous state, and steric effects arise due to finite particle size. Granular gases can exist in a non-equilibrium steady state (NESS) in which energy flows into the system from an injection mechanism and out of the system via dissipation. Studying granular gases allows us to examine the extent to which these systems can be described using the techniques of conventional statistical mechanics.

While there has been a great deal of success describing dilute granular gases using a Chapman-Enskog expansion of the Boltzmann equation \cite{Jenkins1985, Lun1991}, deviations from ordinary gases arise as the density of the gas increases \cite{Goldhirsch-2003-RGF}. For example, granular gases exhibit non-Gaussian velocity distributions \cite{Losert-1999,Falcon1999, Rouyer2000, Wang2009,Hou2008a}, non-equally partitioned kinetic energy (temperature) in mixtures of different particles \cite{Losert-1999, Feitosa2002a, Wildman2002}, and a breakdown of molecular chaos due to velocity correlations \cite{Baxter2007a}. These behaviors, together with simulations in which strong correlations between translational and rotational velocities are observed \cite{Brilliantov2007}, raise the question of how such systems partition energy (or momentum) between rotational and translational modes.  In this Letter, we present an experiment which, remarkably, exhibits equipartition between the rotational and translational energies.

To connect macroscopic state variables (pressure, volume, energy) with microscopic dynamics (translational and rotational particle velocities), we perform experiments on a dense granular gas composed of a single layer of bidisperse disks \cite{Lechenault-2010-EGS, Puckett-2011-LOV}. The particles float on an air hockey table and are driven by impulses from bumpers at the boundaries, producing a fluctuating but stationary state. In contrast with the many previous experiments \cite{Losert-1999, Falcon1999, Rouyer2000, Wildman2002, Feitosa2002a,  Baxter2007a,  Hou2008a,  Wang2009} which inject energy by global agitation, we drive the system via particle-scale bumpers at the boundaries. We find that the distributions of the particle velocities are exponential in character and that they are remarkably unaffected by modifications to the packing density, particle interactions (friction/restitution) or the boundary condition (constant pressure or constant volume).  Because the time-averaged energy of the particles is constant, we are able to identify a gas-like equation of state (EOS) with a van der Waals correction, similar to prior observations in a monodisperse system of vertically-vibrated spheres \cite{Ingale-2010-FOP}. This EOS points to the utility of thermodynamic-like approaches, even for far-from-equilibrium systems.


\begin{figure}[b]
\includegraphics[width=\columnwidth]{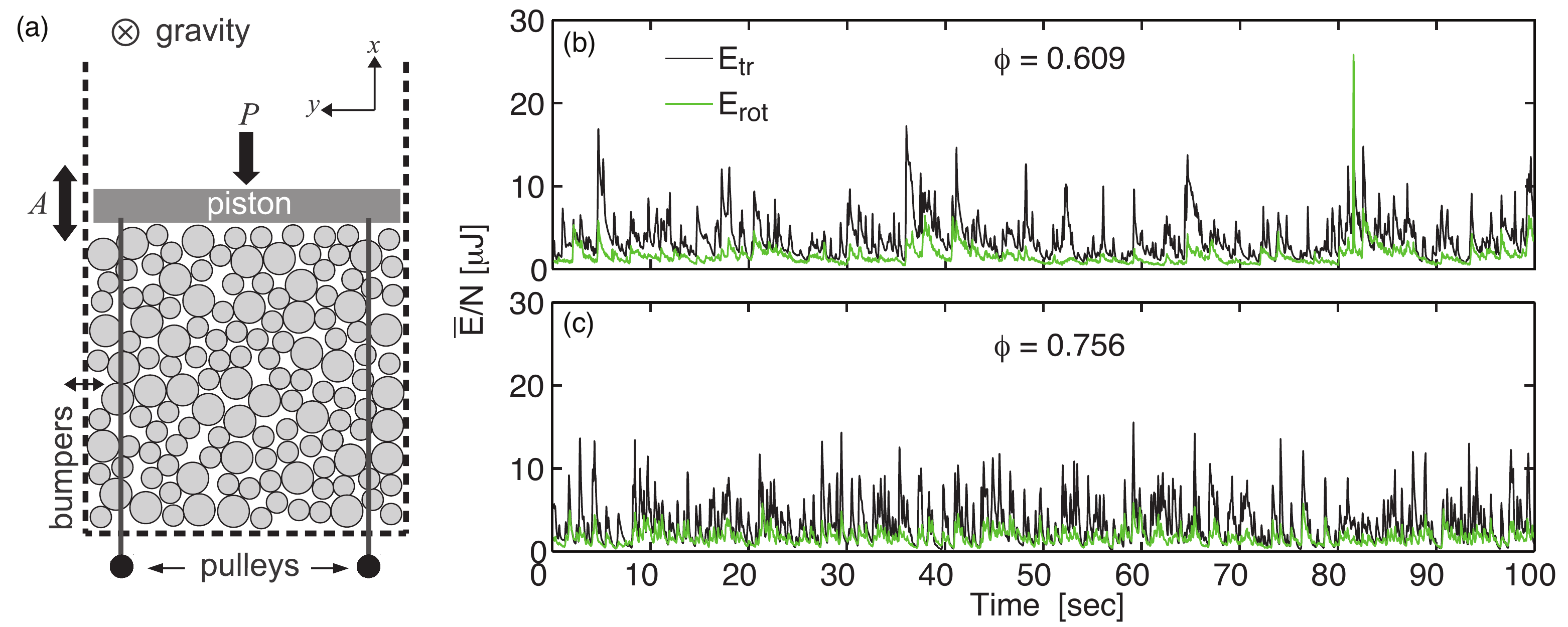}
\caption{Color online. (a) Schematic of experiment, top view, with bumpers at sides. The piston provide CP boundary conditions or be fixed in place to create CV. (b,c) Timeseries of average kinetic energy per particle at two values of $\phi$ (CV boundary conditions).}
\label{fig:apparatus}
\end{figure}

{\bfseries Experiment --- } We perform experiments on bidisperse disks (petri dishes) which float on a horizontal air table (see Fig.~\ref{fig:apparatus}a; further description in \cite{Lechenault-2010-EGS, Puckett-2011-LOV}). We use a high speed camera to track the translational and rotational trajectories at a time-resolution sufficient to resolve individual collisions. The system is driven on three of four sides by impulses from electromagnetic bumpers; random bumpers are triggered every $0.1$~s. As shown in Fig.~\ref{fig:apparatus}b, the injection of kinetic energy is an irregular process, with intermittent strong heating followed by approximately exponential cooling. This irregularity is likely due to transient force chains \cite{Longhi-2002-LFF, Ferguson-2006-SLS} which happen to terminate at a firing bumper and can therefore efficiently transmit impulses. The system is constrained on the fourth side by a piston (mass = 439~g) which also floats on the air table. The position of the piston can be fixed to create a constant volume system (CV), while a constant pressure system (CP) is produced by suspending masses on lines which run from the piston to pulleys on the opposite side of the table. The two-dimensional pressure exerted on the system ranges from $0.01$ N/m $< P < 0.1$~N/m. We primarily investigate amorphous packings with packing density $0.61 < \phi < 0.79$ ($147$ to $186$ particles). These $\phi$ lie below the point at which a jamming transition would occur in a static system \cite{Liu-2010-JTM, vanHecke2010}. The upper limit of the accessible $\phi$ range is set by the pressure at which out-of-plane buckling occurs, while the lower limit is set by the requirement that $\phi$ must be high enough for firing bumpers to contact particles. In addition to amorphous packings, we have also examined crystalline packings in which the system is prepared with all of the small particles on the $+y$ side of the table and all of the large particles on the $-y$ side.


We prepare the system with $N=N_S+N_L$ particles so that the area covered by large and small particles is equal. Large particles have radius $r_L = 41.8$~mm and small particles $r_S = 27.9$~mm; since $(r_L/r_S)^2 = 2$, this requires $N_S = 2 N_L$. For bare petri dish contacts, the coefficient of friction is $\mu = 0.5$ and the coefficient of restitution is $\epsilon=0.33 \pm 0.03$ \cite{Lechenault-2010-EGS}. The friction and restitution coefficients can be adjusted by stretching an elastic band around each of the particles so that the diameter remains unchanged; this results in $\mu' = 0.85$ and $\epsilon' = 0.51 \pm0.07$ \cite{Lechenault-2010-EGS}. (All variables describing the banded particles are denoted with the prime.) The mass $m$ and moment of inertia $I$ for bare dishes are $m_L = 8.13 \pm 0.20$~g and $I_L = 1.3 \pm0.1 \times 10^{4}$~g mm$^2$ and $m_S = 3.52 \pm 0.12$~g and $I_S = 2.5 \pm 0.2 \times 10^3$ g mm$^2$. With the addition of rubber bands, $m$ increases by $0.44$~g, and $I$ increases to $I_S' = 2.8 \pm 0.2 \times10^3$~g mm$^2$ and $I_L' = 1.4 \pm 0.1\times10^4$~g mm$^2$. The moment of inertia is measured by rolling particles down an inclined plane.

Each particle is tagged with a unique nine-dot pattern which allows the identity, position and angular orientation to be determined; in addition, we track the location of the piston to measure the system area $A$. We film the experiment using Phantom V5.2 high speed camera and measure the translational $(v_x, v_y)$ and rotational $(\omega)$ velocities from the trajectory of each particle. A key benefit of unique labels is to eliminate error caused by confusing the identities of two particles \cite{Xu-2004-MEM}. 
Images are taken at a frame rate of $40$~Hz, and a 5-point boxcar average is applied along the trajectories to reduce noise. Sampling at a lower rate causes the velocity distributions to narrow, while taking images at a higher rate leaves the velocity distributions unchanged. Measurements of stationary particles indicate that the $v$ resolution is $\pm 2$~mm/s, while the $\omega$ resolution is $\pm 0.4$~rad/s. The $40$~Hz frame rate allows each measurement run to last $100$~s (4000 frames); $1-4$ such runs are collected for each choice of $\phi$, $P$, the boundary condition (CP, CV) and the particle properties (bare, banded).  We find that stationarity of statistical distributions is achieved within several seconds.


\begin{figure}
\includegraphics[width=\linewidth]{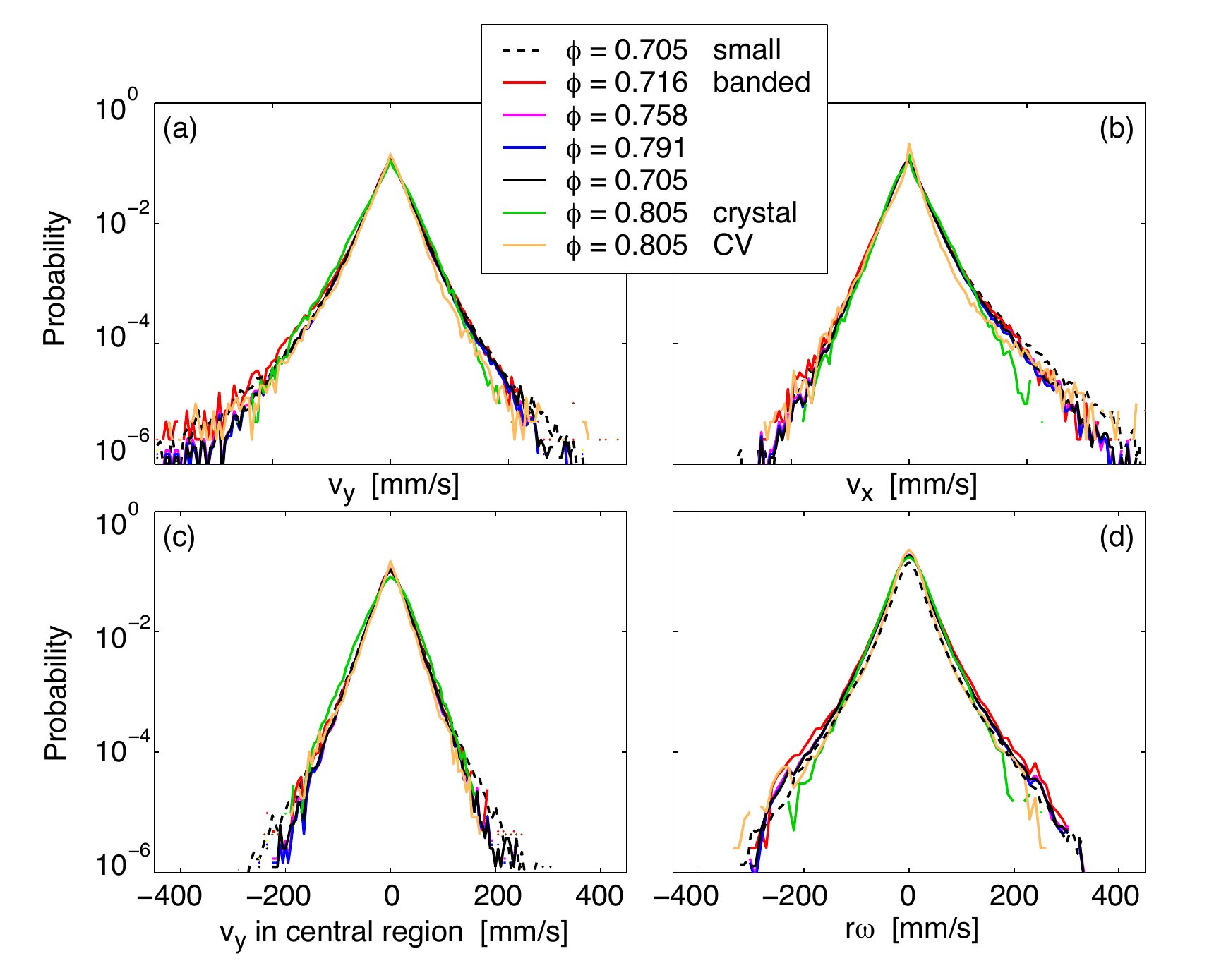}
\caption{Color online. Translational velocity distributions (a) ${\cal P}(v_y)$, with $\langle |v_y| \rangle = 19.9$~mm/s (b)  ${\cal P}(v_x)$, with $\langle |v_x| \rangle = 19.4$~mm/s and (c) ${\cal P}(v_y)$ for particles which are at least $3r_L$ from a boundary, with  $\langle |v_y| \rangle = 19.3$  mm/s. (d) Rescaled rotational velocity distribution ${\cal P}(r \omega)$, with $\langle r|\omega| \rangle = 20.9$~mm/s. Colors indicate conditions of run (large, bare, amorphous, unless otherwise specified.)}
\label{fig:veldist}
\end{figure}

{\bfseries Velocity distributions ---} 
As shown in Fig.~\ref{fig:veldist}a,b, the probability distributions of the translational velocities $v_{x,y}$ are approximately exponential in character, and have the same mean speed $|v_{x,y}| \approx 20$~mm/s. Several observations suggest that the broader-than-exponential tails of ${\cal P}(v)$ are due to high-speed particles near the bumpers. First, ${\cal P}(v_x)$ is asymmetric: the ($-x$) side of the distribution is narrower because there are no bumpers along the piston. Second, if we restrict ${\cal P}(v_y)$ to only those particles more than $3r_L$ from the boundaries, a more exponential form is recovered, shown in Fig.~\ref{fig:veldist}c. The high-velocity tails for the small particles are slightly wider than for the large particles, which suggests that the small particles acquire a slightly larger velocity from the bumpers.

The angular velocity distributions ${\cal P}(\omega)$ are also approximately exponential. In addition, the ${\cal P}(\omega)$ distributions collapse to a single curve when rescaled by the particle radius (see Fig.~\ref{fig:veldist}d), suggesting that the system prefers rolling contacts. Measurements of the relative velocity at each contact confirm this preference. The shapes of all three ${\cal P}(v_x, v_y, \omega)$ distributions are very robust:  switching from CP to CV boundary conditions, varying $\phi$, changing $\mu$ and $\epsilon$, and ordering the system have little effect on any of the velocity distributions. In addition, the mean values of $\langle |v_x| \rangle, \langle |v_y| \rangle$, and $\langle r|\omega| \rangle$ are all similar. As we will examine below, this similarity results from both equipartition and the way mass is distributed within the particles.

\begin{figure}
\includegraphics[width=0.8\linewidth]{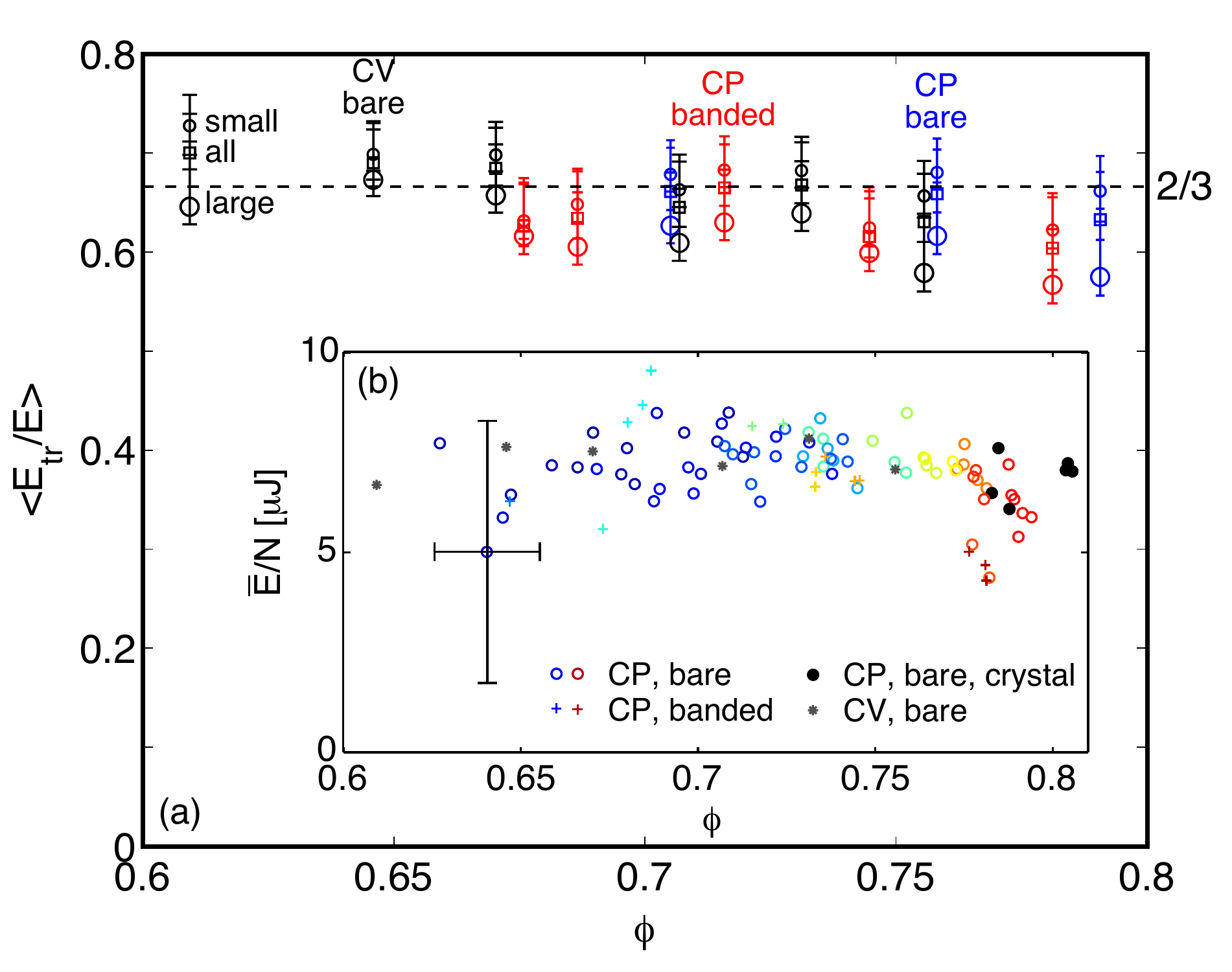} 
\caption{Color online. (a) Time-averaged ratio of the translational kinetic energy to the total kinetic energy $\langle E_\mathrm{tr}/E\rangle$ averaged over $100~$s. Dashed line is 2/3, corresponding to equipartition. (b) Average kinetic energy per particle as a function of $\phi$. Error bars are typical standard deviation.  Each data point indicates the average for $25$~s of data (CP) or $100~$s (CV). Color of symbols indicates $P$ (see Fig.~\ref{fig:EOS} for key).}
\label{fig:equi}
\end{figure}

{\bfseries Equipartition ---}
The time average of total kinetic energy of the system, $\bar{E}=\langle E(t) \rangle$, is the sum of the translational ($\bar{E}_\mathrm{tr}$) and rotational ($\bar{E}_\mathrm{rot}$) energies of all the particles ($i=1..N$) in the system:
\begin{equation}
\label{eq:energy}
\bar{E} = \sum_i \frac{1}{2}m_i \langle v_{i,x}^2 \rangle
   + \sum_i \frac{1}{2}m_i \langle v_{i,y}^2\rangle
   + \sum_i \frac{1}{2} I_i \langle\omega_i^2\rangle.
\end{equation}
The rms velocities, $\langle \cdot \rangle$, are averaged over time. In equilibrium thermal systems, equipartition requires that each degree of freedom (DOF) have the same energy; however, no such requirement exists for non-equilibrium systems. As shown in Fig.~\ref{fig:veldist}, we observe that the large and small particles have the same ${\cal P}(v)$, indicating that translational energy is not partitioned equally between the two species. However, the system {\itshape does} partition the energy equally between the two translational and one rotational DOF: as shown in Fig.~\ref{fig:equi}a, the value of $\langle E_\mathrm{tr}/E \rangle$ is approximately $2/3$. This observation holds for both bare and banded particles and for both CP or CV boundary conditions, although the ratio $\bar{E}_\mathrm{tr}/\bar{E}$ decreases slightly with increasing $\phi$ (or $P$).

In Fig.~\ref{fig:veldist}d we observed that $\langle r_L |\omega_L| \rangle = \langle r_S |\omega_S| \rangle$. However, many different pairs of ${\cal P}(\omega_S)$ and ${\cal P}(\omega_L)$ can satisfy this requirement. An additional constraint is provided by the equipartition of $\bar{E}_\mathrm{tr}$ and $\bar{E}_\mathrm{rot}$:
\begin{equation}
(N_Sm_S+N_Lm_L)\langle v_{x,y}^2 \rangle  = N_S I_S \langle \omega_S^2 \rangle + N_L I_L \langle \omega_L^2 \rangle.
\label{eq:equi}
\end{equation}
The moment of inertia can be written as $I_i = \alpha_i m_i r_i^2$, where $\alpha$ describes how mass is distributed over the particles: $\alpha_S=0.97$, and $\alpha_L=0.94$. Eq.~\ref{eq:equi} can be satisfied for any two values of $\alpha_L$ and $\alpha_S$ when $\langle r_L |\omega_L| \rangle = \langle r_S |\omega_S| \rangle$ and ${\cal P}(v_S) = {\cal P}(v_L)$. The observation that  $\bar{E}_\mathrm{tr}/\bar{E}$ is systematically lower for the large particles follows from the observation that $\alpha_L < \alpha_S$. Because the two ${\cal P}(\omega)$ have the form $e^{-a|\omega_L|/r_L}$ and $e^{-a|\omega_S|/r_S}$, this pair of distributions can fulfill the geometrical preference for rolling contacts for any value of $a$ selected by the system. 
The system is observed to choose $a$ such that there is equipartition of rotational and translational energy at each $\phi$.

{\bfseries Equation of state ---}
In equilibrium statistical mechanics, the total kinetic energy (Eq.~\ref{eq:energy}) corresponds to the temperature: $E = \frac{3}{2} k_B T$. In our experiment, we observe that even though $E(t)$ fluctuates significantly (see Fig.~\ref{fig:apparatus}b), the average per-particle value $\bar{E}/N=\langle E(t) \rangle/N$ is a constant independent of $\phi$, material, boundary condition, and degree of spatial ordering (see Fig.~\ref{fig:equi}b). This value is not universal, but depends on such details as bumper firing frequency, piston inertia, and number of particles. In analogy with a thermal bath at constant $T$, the bumpers at the boundary create a granular gas with constant $\bar{E}$. It is therefore natural to inquire whether a granular version of Boyle's Law ($PV = \mathrm{const.}$) holds.

In two dimensions, the pressure exerted by the piston is given by a force per unit length, so that $P = F/L$ and $PA$ has units of energy. As has been done for experiments on vibrated granular crystals \cite{Ingale-2010-FOP}, we apply a van der Waals-like correction to account for the excluded area. The free area available to the particles is the area under the piston $A(t)$, less the area occupied by the particles and the interstices too small to accommodate another particle. This excluded area is given by $A^* \equiv (N_S \pi r_S^2 + N_L \pi r_L^2)/\phi^*$, where $\phi^*$ is a maximum packing density. In an ideal gas $PA=E$, but in dense systems the compressibility increases due to the relative inaccessibility of nearby configurational states. These steric effects introduce an additional factor $(1+\chi)$ \cite{Luding2001}, where the compressibility factor $\chi$ is proportional to $1+\epsilon$ and to the pair correlation function $g(r)$ evaluated at the particle radius. Including all three considerations, we consider the following EOS:
\begin{equation}
\label{EOS}
P({\bar A} - A^*)= \bar{E}(1+\chi) .
\end{equation}

\begin{figure}
\includegraphics[width=0.9\linewidth]{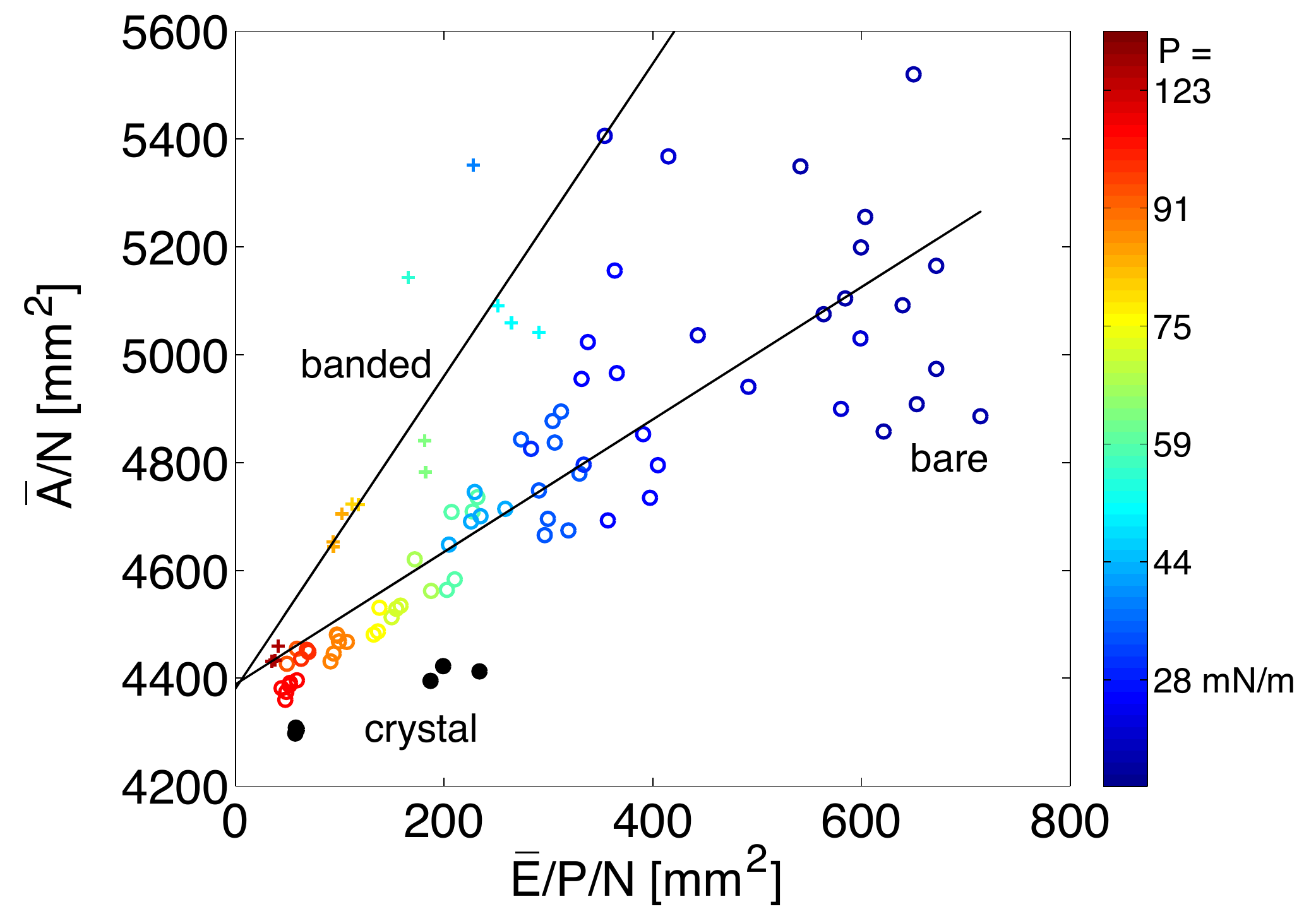} 
\caption{Color online. Data from CP experiments, corresponding to Eq.~\ref{EOS}. Symbol color indicates $P$; symbol shapes as shown in Fig.~\ref{fig:equi}.}
\label{fig:EOS}
\end{figure}

In spite of the large temporal fluctuations of $E(t)$, $\bar{E}= \langle E(t) \rangle$ is constant, as shown in Fig.~\ref{fig:equi}b. We test Eq.~\ref{EOS} by varying $P$ and observing $A(t)$ for a fixed number of either bare or banded particles. We find that there is indeed a linear relationship between ${\bar A}$ and $P$ (Fig.~\ref{fig:EOS}), where the intercept corresponds to the excluded area $A^*$. For $P \lesssim 50$~mN/m, the fluctuations in the area under the piston are large and the system does not behave as though it is in a steady state on the time scale of the experiment. From a least-squares fit weighting each point equally, we determine $\phi^* = 0.79 \pm 0.02$ and $\phi^{'*} =0.79 \pm 0.04$ for the bare and banded particles respectively. For comparison, previously measured \cite{Lechenault-2010-EGS} values of static random loose packing estimate $\phi_\mathrm{RLP} = 0.81$ for both bare and banded particles. For the crystalline system, $\phi^*_\mathrm{cryst}=0.81 \pm 0.02$, which is also less dense than the hexagonally close packed value $\phi_\mathrm{HCP} = 0.91$. Therefore, the measured $\phi^*$ do not correspond to the van der Waals-like maximally dense static state (this would corresond to either random-close-pacing $\phi_{RCP}$ or $\phi_{HCP}$), but to some kinetically-determined value.

The slope of the equations of state plotted in Fig.~\ref{fig:EOS} correspond to $1+\chi$; their deviation from a value of $1$ corresponds a measurement of the compressibility.  For bare particles, $1+\chi = 1.3\pm0.1$, for banded particles  $1+ \chi' = 3.2\pm0.8$, and for crystalline bare particles $1+\chi_\mathrm{cryst} = 0.7\pm0.1$. Although $\chi$ is predicted to increase as the system becomes denser \cite{Luding2001}, we instead find that $\chi$ is approximately constant for all the values of $P$ that we examined. This result is unexpected, especially given that experiments on a single layer of vertically-shake spheres \cite{Ingale-2010-FOP} indicate that  $\chi$ rises with increasing $P$ as expected.


{\bfseries Conclusions \& Discussion ---} 
We have tracked the translational and rotational motion of a dense granular gas of bidisperse particles driven by bumpers at the boundary of a horizontal air table. The states generated have constant $P$, $\bar A$, and $\bar E$, which motivates treating the system within the framework of statistical physics. The velocity distributions are approximately exponential and are independent of the packing density, particle properties, and degree of spatial ordering. The exponential character of these distributions is consistent with both horizontal quasi-2D free cooling experiments \cite{Losert-1999} and quasi-2D vibration experiments in microgravity \cite{Falcon1999} and \cite{Hou2008a}. Our system appears to experimentally realize a state in which rare, high-energy injections produce a well-mixed steady state \cite{Ben-Naim2005}, although simulations of this driving mechanism find power law tails in the velocity distributions, while we observed nearly-exponential tails. Although energy is not partitioned equally between the large and small particles, we do observe equipartition between the two translational and one rotational DOF. Finally, the system exhibits a van der Waals-like equation of state. These findings indicate that while granular gases  can exhibit some properties of equilibrium statistical mechanics (equipartition of rotational and translational energy, Boyle's Law), they do not necessarily do so (the velocity distributions are not Gaussian and energy is not partitioned equally between the large and small particles). The observation of a thermodynamic-like EOS for such a small system provides encouragement that similar descriptions may be possible in the many boundary-drive granular systems where only a mesoscopic portion of the system, typically $<10$ particle diameters within shear bands, is set into motion. 

We find that the translational velocity distributions are the same for both the large and small particles; consequently, energy is not partitioned equally between the two particle species. Other experiments on ensembles of particles made from different materials \cite{Losert-1999, Feitosa2002a, Wildman2002} indicate that energy is not partitioned equally between particles with different properties. The absence of equipartition amongst particles with different restitution coefficients has been explained within the context of Chapman-Enskog expansions \cite{Garzo1999}. In addition, simulations \cite{McNamara1998, Herbst2005} predict that the way energy is distributed between rotational and translational modes  depends on $\mu$ and $\epsilon$; however, for the values used here, equipartition of $\bar{E}_\mathrm{tr}$ and $\bar{E}_\mathrm{rot}$ is not predicted. 

While we observe equipartition between the rotational and translational degrees of freedom, it is notable that no such equipartition is observed in dilute granular gases with fully 3D motions \cite{Feitosa2011}; this raises the question of whether the density or the dimensionality is the key difference which promotes equipartition. Additional constraints arise due to density: particles most easily translate in concert with their neighbors and rotational motion is governed by interparticle friction. That ${\cal P}(v_{x,y})$ are the same for both the large and the
small particles in our 2D system indicates that translational motion
requires cooperation. In addition, that ${\cal P}(v_{r} \omega)$ are
the same for both the large and small particles indicates a
preference for rolling contacts. The latter constraint does not apply in 3D because translational motion can occur along the axes of rotational motion. Once the preference for rolling contacts is fulfilled, our 2D system still must select how to distribute kinetic energy amongst the rotational and translational degrees of freedom. Unexpectedly, our findings suggest that equipartition, while not guaranteed for a non-equilibrium system, is used to select how the rotational and translational energy is distributed.


{\bf Acknowledgements --- } We are grateful for support from an NSF Career Award DMR-0644743 (KD) and from FOM (KN). We would also like to thank F. Lechenault, J. Puckett and E. Owens for their instrumentation and particle-tracking acumen and M. Shattuck and N. Menon for helpful discussions and sharing unpublished results on related systems.

\bibliographystyle{apsrev}

\end{document}